\documentclass[mathleft
]{an}
\usepackage{graphicx}
\usepackage{times}
\begin{document}

\Pagespan{1}{}
\Yearpublication{2007}%
\Yearsubmission{2007}%
\Month{11}%
\Volume{328}%
\Issue{10}%

\title{The differential rotation of $\varepsilon$ Eri from
MOST data\thanks{Based on data from the MOST satellite, a Canadian Space Agency mission,
jointly operated by Dynacon Inc., the University of Toronto Institute for
Aerospace Studies and the University of British Columbia, with the assistance
of the University of Vienna.}}

\author{H.-E. Fr\"ohlich\inst{}\thanks{Corresponding author:
  {HEFroehlich@aip.de}\newline}
}
\titlerunning{Differential rotation of $\varepsilon$ Eri}
\authorrunning{H.-E. Fr\"ohlich}
\institute{
Astrophysikalisches Institut Potsdam, An der Sternwarte 16, 
D-14482 Potsdam, Germany
}

\received{2007 Nov 2}
\accepted{2007 Nov 5}
\publonline{2007}


\abstract{%
  From high-precision MOST photometry spanning 35 days the existence of two
  spots rotating with slightly differing periods is confirmed. 
  From the marginal probability distribution of the derived 
  differential rotation parameter $k$ its {\it expectation} value as well as
  confidence limits are computed directly from the data. 
  The result depends on the assumed range in inclination $i$, not on the shape
  of the prior distributions. 
  Two cases have been considered: (a) The priors for angles, inclination $i$ of
  the star and spot latitudes $\beta_{1,2}$, are assumed to be constant 
  over $i$, $\beta_1$, and $\beta_2$. (b) The priors are assumed to be constant
  over $\cos i$, $\sin\beta_1$, and $\sin\beta_2$. In both cases the full range
  of inclination is considered: $0^\circ\le i\le 90^\circ$. Scale-free parameters, 
  i.\,e.  periods and spot areas (in case of small spots)  are taken logarithmically. 
  Irrespective of the shape of the prior, $k$ is restricted to $0.03\le k\le 0.10$ 
  (one-$\sigma$ limits). The inclination $i$ of the star is photometrically
  ill-defined. 
}
\maketitle

\section{Introduction}

In late-type stars with their deep convection zones non-uni\-form rotation is
driven by the action of the Coriolis force on that convective turbulence 
(cf. Kitchatinov \& R\"udiger 1993). Now quantitative models of 
differential rotation for the Sun and solar-like stars are available
(R\"udiger \& Kitcha\-tinov  2005; 
K\"uker \& R\"udiger 2005; K\"uker [these proceedings]) which should be compared with real stars.

The outcome of these theoretical efforts is a lapping time which is
for a given main-sequence star nearly independent of its rotation period (in 
the case of the Sun, a G2 dwarf, roughly 100 days). 

Precision photometry of a spotted star with spots differing in latitude 
allows a direct measurement of the differential rotation 
parameter $k$. It parameterizes the surface rotation. 
With $P_{\rm eq}$ denoting the equatorial rotation period that at
latitude $\beta$ is $P_\beta = P_{\rm eq}/(1-k\,\sin^2\beta)$.

Here the MOST data (Croll et al. 2006; Croll 2006) 
of the star $\varepsilon$ Eri have been reanalyzed
in a Bayesian framework. 
The motivation was to get realistic error estimates
by computing $k$'s marginal distribution and to find out, how it depends
on the chosen prior. 
In the essence a Bayesian approach explores the whole 
likelihood mountain in the $N$-dimensional parameter space. 
It looks not only to the most probable set of parameter
values, but provides {\em expectation\/} values. Moreover, from a marginal
distribution reliable confidence limits can be easily derived. 

The nearby K2 dwarf is young ($\le$ 1 Gyr), rotates
twice as fast as the Sun, and shows strong chromospheric activity 
(cf. Biazzo et al. 2007). 

\section{The MOST data}

\begin{figure}
\includegraphics[width=82mm]{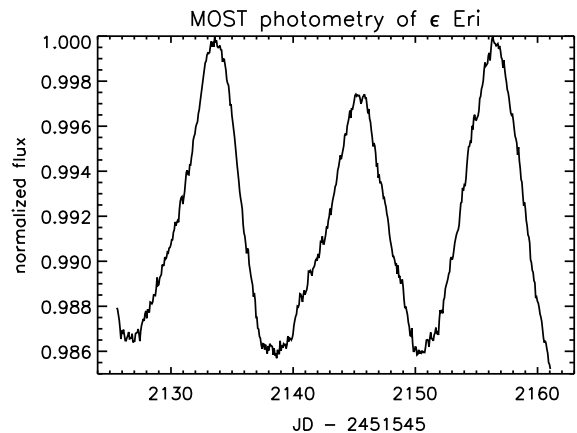}
\caption{The $\varepsilon$ Eri light curve with an obvious trend removed.}
\label{fig01}
\end{figure}

The MOST photometric satellite (Walker et al. 2003) observed three 
consecutive rotations of $\varepsilon$ Eri in 2005. 
The data consist of 492 data points, one point per orbit of the
satellite. The point-to-point precision of the data is 50 ppm rms (i.e. $\pm$
0.000\,05 mag). The rectified light curve (Fig. \ref{fig01}) spanning 35 days 
shows an overall variation of a hundredth of a magnitude.

\section{A Bayesian data analysis}

In the case of two circular spots the light curve is determined by nine
parameters: two periods, two epochs, two 
latitudes, two areas, and the inclination of the star. 
The parameters specifying limb darkening and spot rest
intensity are assumed to be given (cf. Croll et al. 2006). 

There are nuisance parameters: an offset in the
photometric zero point and a long term trend in the data. With these two
parameters introduced shifting the
light curve vertically and even adding a trend does not alter the results. 
The flux error is considered a Gaussian with unknown
variance. Integrating away that error $\sigma$, assuming
Jeffreys $1/\sigma$ prior, sounds strange, but perhaps it is not so bad an
assumption and it leads to quite a simple formula for the likelihood. 
The likelihood function used is therefore a {\em mean\/} likelihood with respect 
to these three parameters. One can say the 
method itself determines offset, slope and measurement error.

The result, especially the star's inclination $i$, depends on the prior distribution functions. 
In the following it is assumed that rotational frequencies as well as spot
areas (if small compared with the star), i.\,e. parameters missing a characteristic scale, 
have constant priors if taken logarithmically. 
Two cases are considered: In Case A the prior is assumed constant over the 
inclination $i$ and the spot latitudes $\beta_1$, and $\beta_2$ as in the Croll (2006) paper. 
In Case B it is assumed constant over
$\cos i$, $\sin\beta_1$, and $\sin\beta_2$, respectively.
An inclination  prior constant over $\cos i$ means that nothing is a priori known about the
orientation of the rotational axis. A latitude prior constant over $\sin\beta$
takes into account that there are more possibilities to locate by chance
a spot near the equator than near the poles. 

In order to explore the likelihood mountain over 
a nine-dimensional parameter space the Markov chain Monte Car\-lo (MCMC) 
method (cf. Press et al., 2007) has been applied.
The results should be therefore best compared with the analysis of the
MOST data by Croll (2006), who has employed that MCMC technique. 

A set of 64 Markov chains was generated. 
Each chain has performed $10^7$ steps. 
After a burn-in period of 1000 steps every 1000th successful step has been
recorded to suppress the correlation between successive steps. 
For modeling light curves Budding's star-spot model (1977) has been used.

\section{Results}

\begin{figure}
\includegraphics[width=82mm]{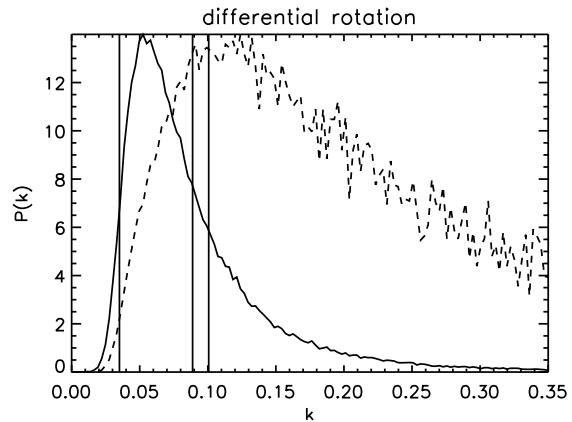}
\caption{The marginal distribution for the differential rotation 
parameter $k$ (Case B). Vertical lines denote expectation value and the 68 per cent
confidence region. Additionally to the marginalized distribution 
the run of {\em mean\/} likelihood (dashed) is given, providing an impression
how the goodness of fit varies with $k$. The mismatch hints at the
non-Gaussianity of the likelihood mountain (cf. Lewis \& Bridle 2002).}
\label{fig02}
\end{figure}

\begin{figure}
\includegraphics[width=82mm]{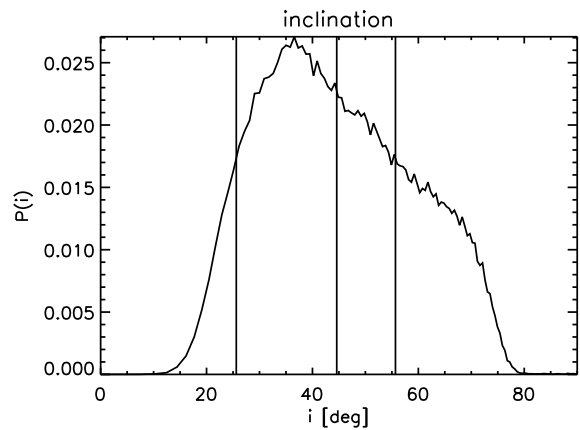}
\caption{The marginal distribution for the inclination $i$ (Case B). Vertical
lines denote expectation value and the 68 per cent confidence region. The
high-$i$ solution is suppressed by the chosen prior.}
\label{fig03}
\end{figure}

\begin{figure}
\includegraphics[width=82mm]{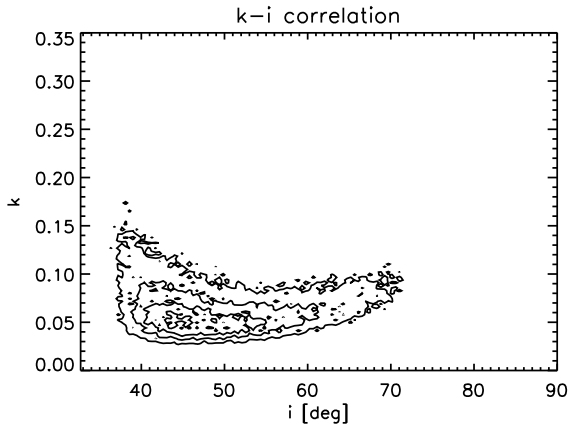}
\caption{The parameters of differential rotation $k$ and inclination $i$ are
somewhat correlated (Case B).}
\label{fig04}
\end{figure}

Expectation values and modes as well as one-$\sigma$ confidence limits 
are presented in Table~\ref{tab01}, augmented by the solution of
Croll (2006; his ``Wide Prior'' Case). 
The Case-B probability distributions for the parameters $k$ and $i$ are shown in
Figs.~\ref{fig02} and \ref{fig03}, respectively. Especially the inclination
parameter $i$ proves ill-defined by photometry alone. 
There is no strong correlation between $i$ and $k$ (Fig.~\ref{fig04}).

The differences between the two cases A and B are mar\-gin\-al, i.\,e.  
the outcome is rather insensitive to the {\em shape\/} of the prior. 
Most important
is the restriction of the inclination $i$ to values between
$15^\circ$ and $40^\circ$ in the ``Wide Prior'' Case of the Croll MCMC analysis. 
(The marginalized likelihood of $i$, the last plot of his Fig.~3,
indicates that inclinations beyond $40^\circ$ are not ruled out.)
Here the whole range, $0^\circ\le i\le 90^\circ$, is taken into account. 
It includes a second peak in the likelihood
mountain: $i\approx 72^\circ, \beta_1\approx 61^\circ, 
\beta_2\approx 73^\circ$. 
This high-$i$ solution is even more probable, by a factor of
4.6, then the {\em best\/} low-$i$ solution ($i\approx 24^\circ, \beta_1\approx
14^\circ, \beta_2\approx 25^\circ$). 

Contrary to the Croll solution the second spot is always visible.

\begin{table*}
\centering
\caption{Results: Given are the {\em expectation\/} values, the modes (Case B) 
and $1 \sigma$ errors. 
Periods $P$ are in days, epochs are with regard to $E$ = HJD - 2,451,545.
Note that the Case-A $i$ and $\beta_1$ are both bi-modal. To give only one
interval, for once the 90\% confidence region is chosen.}
\label{tab01}
\begin{tabular}{l@{~}ll|*{3}{r@{}l}|*{1}{r@{}l}}\hline\\ [-9pt]
\multicolumn{3}{l}{parameter} & \multicolumn{2}{|c}{Case A} & \multicolumn{4}{c|}{Case B} & \multicolumn{2}{c}{Croll (2006)}\\
&&              & \multicolumn{2}{c}{mean}   & \multicolumn{2}{c}{mean}   & \multicolumn{2}{c|}{mode}   & \multicolumn{2}{c}{mode}\\ 
&&&\multicolumn{6}{c|}{$0^\circ ~~~~~ \le ~~~ i ~~~ \le ~~~~~ 90^\circ$}&\multicolumn{2}{c}{$15^\circ\le i\le 40^\circ$}\\
\hline
&\\[-8pt]
\multicolumn{2}{l}{differential rotation} &$k$     & $0.088$&$^{+0.011}_{-0.055}$                & $0.089$&$^{+0.012}_{-0.054}$ &
$0$&$.053$      &
$0.058$&$^{+0.084}_{-0.018}$\\[4pt]
\multicolumn{2}{l}{inclination}    &$i$            & $46^\circ\hspace{-3pt}.6$&$^{+26}_{-23}$    & $44^\circ\hspace{-3pt}.6$&$^{+11}_{-19}$&
$36$&$^\circ$   &
$25^\circ\hspace{-3pt}.9$&$^{+12.6}_{-6.7}$\\[4pt]
1st &latitude   &$\beta_1$   & $35^\circ\hspace{-3pt}.1$&$^{+25}_{-23}$       & $33^\circ\hspace{-3pt}.0$&$^{+9.1}_{-19}$       &
$25$&$^\circ$   &
$16^\circ\hspace{-3pt}.5$&$^{+7.0}_{-5.0}$\\[4pt]
2nd &latitude   &$\beta_2$   & $51^\circ\hspace{-3pt}.2$&$^{+21}_{-12}$       & $48^\circ\hspace{-3pt}.8$&$^{+18}_{-14}$        &
$48$&$^\circ$   &
$24^\circ\hspace{-3pt}.8$&$^{+15.1}_{-3.5}$\\[4pt]
1st &radius     &$\gamma_1$  & $5^\circ\hspace{-3pt}.71$&$^{+0.25}_{-1.1}$    & $5^\circ\hspace{-3pt}.62$&$^{+0.20}_{-0.98}$    &
$4$&$.9^\circ$  &
$5^\circ\hspace{-3pt}.3$&$^{+2.1}_{-0.2}$\\[4pt]
2nd &radius     &$\gamma_2$  & $7^\circ\hspace{-3pt}.78$&$^{+0.15}_{-1.6}$    & $7^\circ\hspace{-3pt}.49$&$^{+0.07}_{-1.3}$     &
$6$&$.6^\circ$  &
$6^\circ\hspace{-3pt}.8$&$^{+1.4}_{-0.3}$\\[4pt]
1st &period     &$P_1$       & $11.348$&$^{+0.037}_{-0.036}$                  & $11.349$&$^{+0.037}_{-0.034}$    &
$11$&$.35$      &
$11.35$&$^{+0.03}_{-0.03}$\\[4pt]
2nd &period     &$P_2$       & $11.553$&$^{+0.020}_{-0.020}$                  & $11.554$&$^{+0.019}_{-0.020}$    &
$11$&$.555$     &
$11.55$&$^{+0.02}_{-0.02}$\\[4pt]
1st &epoch      &$E_1$       & $2130.43$&$^{+0.20}_{-0.21}$                   & $2130.41$&$^{+0.19}_{-0.22}$     &
$2130$&$.37$    &
$2130.43$&$^{+0.20}_{-0.21}$\\[4pt]
2nd &epoch      &$E_2$       & $2126.47$&$^{+0.11}_{-0.11}$                   & $2126.46$&$^{+0.11}_{-0.12}$     &
$2126$&$.46$    &
$2126.47$&$^{+0.11}_{-0.12}$\\ [2pt]
\hline
\end{tabular}
\end{table*}

By fitting spectroscopic measurements (2000/2001) to a spot/plage model 
Biazzo et al. (2007) find two spots but larger than ours. 
By the way, this could be due to the lower temperature contrast between spot and photosphere
these authors have found. Perhaps the spots are persistent. 
The stated spot latitudes, $\beta_1\approx 21^\circ$ and $\beta_2\approx 48^\circ$, 
are within the uncertainties of our estimate.

Equatorial rotational period ($P_{\rm eq} = 11.2$ d) and radius ($0.72 R_\odot$) 
of $\varepsilon$ Eri are well known. So in
principle spectroscopic determinations of the projected rotational velocity 
may restrict the inclination $i$. 
Unfortunately, $\varepsilon$ Eri rotates slowly. 
Saar \& Osten (1997) find $v\,\sin i \approx 1.7\pm 0.3$ km/s.
This leads to a $\sin i = 0.5\pm 0.3$ hinting at the low-$i$ solution, 
but is too uncertain as to constrain the range of $i$
values very much. 
A recent compilation by Valenti \& Fischer (2005) gives 
$v\,\sin i \approx 2.4\pm 0.5$ km/s. This favours an 
inclination close to $50^\circ$.

One should note that the planetary companion (Hatzes
et al. 2000) as well as a 130 AU dust ring show both an inclination distinctly below
$30^\circ$, namely 
$26^\circ\hspace{-3pt}.2$ (Benedict et al. 2006) and $\approx 25^\circ$
(Greaves et al. 2005), respectively.

\section{Conclusions}

As expected, with $0.03\le k\le 0.10$, 
the estimated differential rotation parameter $k$ proves smaller than 
that of the Sun ($k_\odot\approx 0.2$).	
The horizontal shear is $0.017\le\delta\Omega\le 0.056$ rad/d, the lapping
time 130 d.
An independent estimate of the star's inclination would exclude either 
the low-$i$ solution or the high-$i$ one and would help to constrain $k$ even more.

\bigskip
Acknowledgments: The author thanks the MOST team for the $\varepsilon$ Eri
data and W.W. Weiss from Vienna for valuable discussions.


\begin{thebibliography}{}
\bibitem{} Benedict, G.F., McArthur, B., Gatewood, G., et al.: 2006, AJ~132, 2206
\bibitem{} Biazzo, K., Frasca, A., Henry, G.W., Catalano, S., Marilli, E.: 2007, ApJ~656, 474
\bibitem{} Budding, E.: 1977, Ap\&SS~48, 207
\bibitem{} Croll, B., Walker, G.A.H., Kuschnig, R., et al.: 2006, ApJ~648, 607
\bibitem{} Croll, B.: 2006, PASP~118, 1351
\bibitem{} Greaves, J.S., Holland, W.S., Wyatt, M.C., et al.: 2005, ApJ~619, L187
\bibitem{} Hatzes, A.P., Cochran, W.D., McArthur, B., et al.: 2000 ApJ~544, L145
\bibitem{} Kitchatinov, L.L., R\"udiger, G.: 1993, A\&A~276, 96
\bibitem{} K\"uker, M., R\"udiger, G.: 2005, AN~326, 265
\bibitem{} K\"uker, M.: 2007, this volume
\bibitem{} Lewis, A., Bridle, S.: 2002, PhRvD 66, 103511
\bibitem{} R\"udiger, G., Kitchatinov, L.L.: 2005, AN~326, 379
\bibitem{} Press, W.H., Teukolsky, S.A., Vetterling, W.T., Flannery, B.P.: 
           2007, {\it Numerical Recipes}, 3rd Edition, Cambridge University Press
\bibitem{} Saar, S.H., Osten, R.A.: 1997, MNRAS~284, 803
\bibitem{} Valenti, J.A., Fischer, D.A.: 2005, ApJS~159, 141
\bibitem{} Walker, G., Matthews, J., Kuschnig, R., et al.: 2003, PASP~115, 1023
\end{thebibliography}
\end{document}